\documentclass[prl,twocolumn,showpacs]{revtex4}
\usepackage{epsfig}
\begin{document}
%
%
%
\title{
Occupation Time Statistics in the Quenched Trap Model
}

\author{S. Burov, E. Barkai\\
Department of Physics,
Bar Ilan University, Ramat-Gan 52900 Israel
}
\begin{abstract}
{ 
We investigate the distribution of occupation times
for a particle undergoing a random walk among random energy traps
and in the presence of  a deterministic potential field
$U^{{\rm det}}(x)$. 
 When the distribution of energy traps is exponential
with a width $T_g$  we find that the occupation
time statistics behaves according to (i) the canonical 
Boltzmann theory when $T>T_g$, (ii) 
while for $T<T_g$  they are distributed according to the Lamperti
distribution  
with the asymmetry of the distribution 
determined by the Boltzmann
factor $\exp( -U^{{\rm det}}(x)/T_g)$ with     
$T_g$ and not $T$ being the effective temperature.    
We explain how our results describe occupation
times in other systems with quenched disorder, when the underlying
partition function of the problem
is a random variable distributed according to L\'evy statistics. 
}
\end{abstract}

\maketitle

 Consider a one dimensional Brownian motion,
in a binding deterministic potential
field $U^{{\rm det}}(x)$.
Following the trajectory of an individual
particle one may determine the occupation time 
$t^{{\rm Occ}}$, which is the time the particle
spends  in 
the domain $x_1<x<x_2$. 
In the limit of long measurement
times $t$ the occupation fraction $\overline{p} \equiv t^{{\rm Occ}}/t$ is given
by Boltzmann statistics, assuming that the dynamics is ergodic  
\begin{equation}
\overline{p}  \to{ \int_{x_1} ^{x_2} e^{ - { U^{{\rm det}} \left( x \right) \over T}} {\rm d} x \over Z},
\label{eqInt01}
\end{equation}
where  
$Z=\int_{-\infty} ^{\infty} \exp[ -  U^{{\rm det}} \left( x \right)
/ T] {\rm d} x$ 
is the normalizing partition function 
and $T$ is the temperature.
We see that the occupation fraction $\overline{p}$ does not
depend on the details of the dynamics, for example
on the diffusion constant or the initial conditions,
which is of-course the strength
and the  generality of statistical mechanics.  
Majumdar and Comtet showed that 
when the potential energy is random the occupation
fraction may exhibit large fluctuations from one
sample of disorder to another, in particular in
 \cite{Majumdar} the Sinai model was investigated. 
Generally, for disordered systems an important  question is what
is the distribution of the occupation fraction and its
relation to the underlying disorder. 

Here we investigate occupation time statistics using the
well known trap model \cite{WEB,DEAN}. 
Briefly the model introduced in the seventies,
describes a random walk among traps 
with a random depths of energy traps, with a density
of states which is exponentially distributed with a width 
$T_g$ (see details below).
Such density of states leads to anomalous diffusion \cite{Monthus,Bertin1},
and aging \cite{WEB,Monthus,Bertin,Rinn} when $T<T_g$.
The model, in fact a family of models,
was used to describe dynamics of many systems: transport
of electrons in amorphous materials \cite{Cohen,Silver,Kastner}, 
single molecule pulling experiments \cite{Julio},
rheology of soft matter \cite{Sollich1}
e.g. emulsions, relaxation in glasses \cite{WEB,DEAN,Bercu}
and green fluorescent
protein dynamics \cite{Bardou} to name a
few. Recently the aging scenario of the trap model \cite{WEB,DEAN}, 
was justified rigorously using
the random energy model as the starting point \cite{Ben}.

Bouchuad \cite{WEB} introduced the concept of weak ergodicity breaking,
in the context of the trap model,
which implies the breakdown of Boltzmann's statistics  when $T<T_g$.
The goal of this paper is to define and investigate
the rather strong deviations from Boltzmann's
statistics in the quenched trap model.
In particular we investigate the distribution of the
occupation times for the quenched trap model
quantifying the deviations from the standard canonical 
theory 
Eq. (\ref{eqInt01}).
At the end of the paper we will  show how the main features
of  our theory
can
describe occupation times in several other models of quenched disorder,
proving some generality of our results beyond the trap model.

{\em Quenched Trap Model} 
We consider a particle undergoing a one dimensional random
walk on a quenched random energy landscape on a lattice. 
Lattice points are on $x=0,a,2 a, \cdots,L$ where $a$ 
is the lattice spacing. On each lattice point a random energy
$E_x$ is assigned, which is minus the energy of the particle
on site $x$, so $E_x>0$ is the depth of a trap on site $x$.
The traps energies $\{E_x\}$ are independent identically
distributed random variables, with a common probability density function
(PDF)
$\rho(E)= (1 / T_g)\exp(-E/T_g)$. Due to an interaction
with a heat bath the particle may escape site 
$x$ and jump to one of its nearest neighbors. The average time
it takes the particle to escape from site $x$ is given
by Arrehhius law $\tau_x = \exp(E_x/T)$.
Notice that small changes in $E_x$ leads to an exponential  shift
in $\tau_x$. In particular it is easy to show that the
PDF of the waiting times is
\begin{equation}
\psi(\tau) = {T \over T_g}  \tau^{ - (1 + {T\over T_g})} \ \ \ \ \tau\ge 1
\end{equation} 
so when $T<T_g$ the average waiting time diverges.

An additional bias is applied to the system.
For example in transport processes this is a driving field, e.g. an external
electric field (and see more details below). 
Let $q_x$ $(1-q_x)$ be the probability of jumping left 
(right) from site $x$ respectively. The master equation
for the population on site $x$, $P_x$ is
\begin{equation}
{ {\rm d} P_x \over {\rm d} t} = - { 1\over \tau_x}P_x +{ q_{x+1} \over \tau_{x+1} }  P_{x+1} + { 1 - q_{x-1} \over \tau_{x-1} } P_{x-1} . 
\label{eq02}
\end{equation}
For non biased random walks $q_x = 1/2$ while for uniformly biased
random walks $q_x\ne 1/2 $ is a constant. The boundary conditions
are reflecting $q_0=0$ and $q_L=1$ though our main results are valid
also for periodic boundary conditions. 

 The local bias $q_x$ is controlled by a deterministic potential field $U_x ^{{\rm det}}$, 
which in
some cases is controlled by the experimentalist.
It is usually assumed that detailed balance conditions holds so that
the dynamics of the populations  reaches thermal equilibrium
described by Boltzmann's canonical ensemble. For the trap model
this well known condition leads to 
\begin{equation}
{ q_x \over 1 - q_{x-1} } = \exp \left[ - {\left( U^{{\rm det}} _{x-1} - U ^{{\rm det}} _{x} \right) \over T } \right].
\label{eq}
\end{equation}   
For example if a constant driving force field ${\cal F}$ acts on the
system $q_x=1/ [ 1 + \exp({\cal F} a / T)]$ \cite{Bertin}.

 We consider a single realization of disorder in the thermodynamic
limit where the measurement time $t \to \infty$ before the system size
is made large. One can show that the equilibrium of populations
is described by Boltzmann statistics, which is not surprising since
we used the detailed balance condition. The total time the
particle spends in the domain $x_1\le x \le L$ is the occupation
time $t^{{\rm Occ}}$. This domain is called the observation
domain.   Assuming that the process is 
ergodic  we have the occupation fraction for a single
disordered system
\begin{equation}
\overline{p} = { t^{{\rm Occ}} \over t} \to {Z^{{\rm O}} \over Z^{{\rm O}} + Z^{{\rm NO}} } 
\label{eq04}
\end{equation}
where 
\begin{equation}
Z^{{\rm O}}= \sum_{x=x_1} ^L \exp\left[ - {\left(U^{{\rm det}}_x - E_x \right)\over  T }\right]
\label{eq04a}
\end{equation}
is the partition function of the part of the system under observation
and 
$Z^{{\rm NO}} = \sum_{x=0} ^{x_1 - a} \exp[- ( U^{{\rm det}} _x - E_x )/T ]$ is the partition function of the rest 
of the system. The occupation fraction is a random variable which
varies from one system to the other, the goal of this manuscript
is to calculate its distribution. However first four comments
are in place. (i) If we have only a single realization of disorder
the occupation fraction is given by Boltzmann statistics, the
question then is whether the occupation fraction a self averaging
quantity. Namely is it reproducible in a second experiment
when a different realization of disorder is investigated. 
More generally we have in mind the situation where
one investigates many realizations of disorder,
for each the occupation fraction is random and
hence one may construct its distribution, this case
corresponds in principle to single molecule experiments where one may track
independently a large number
of  individual molecules
each one interacting with a unique random environment
\cite{BarkaiRev}.
(ii)  We have assumed that for a single disordered system the
dynamics is ergodic. This is so since we are considering
the thermodynamic limit where the measurement time $t \to \infty$
first while maintaining a finite size of the system. For
a finite system we always have a finite $E_{{\rm max}}$ which
is the maximum of the random  energies $\{E_{x} \}$, and hence there
is always a long time $t$ which is much larger than $\exp(E_{{\rm max}}/T)$
after which the process is behaving according to the ergodic principle. 
(iii) The occupation time in
Eq. (\ref{eq04}) describes rather generally the occupation time
of a particle in a random energy landscape and
is not unique to the specific dynamics
of the quenched trap model. 
For example we could add random barriers to the dynamics
of the  model which would not alter its equilibrium.  
(iv) Previous work \cite{Bel} considered the occupation times
of the continuous time random
walk (CTRW) model (annealed model),  unlike
the quenched trap model in the CTRW model ergodicity is broken
and the system is not spatially disordered. 

From Eq. (\ref{eq04}) we see that the distribution of
the occupation fraction $\overline{p}$ is obtained
in principle from the distributions of two independent
random partition functions $Z^{{\rm O}}$ and $Z^{{\rm NO}}$.
Let $G_{Z^{{\rm O}}}(z)$ and
$G_{Z^{{\rm NO}}}(z)$ be the PDFs
of $Z^{{\rm O}}$ and $Z^{{\rm NO}}$ respectively. Then the PDF of
the occupation
fraction $f(\overline{p})$ 
is found using Eq. 
(\ref{eq04})
\begin{equation}
f\left( \overline{p} \right) = \int_0 ^\infty {\rm d} z z
G_{Z^{{\rm NO}} } \left[ \left( 1 - \overline{p} \right) z \right] G_{Z^{{\rm O}} } \left( \overline{p} z \right).
\label{eq05}
\end{equation}
 We now consider the problem of finding $G_{Z^{{\rm O}}}(z)$. 

If the deterministic part of the field 
$U^{{\rm det}}_x$  is a constant, Eq. (\ref{eq04a}) shows that
$Z^{{\rm O}}$ is a  sum
of independent identically distributed random variables, and
then Gauss--L\'evy limit theorems apply. In contrast, when
$U^{{\rm det}}_x$ is not a constant then we are dealing with the
problem of summation of non-identically distributed random variables  
and hence in what follows we modify
the familiar limit theorems for the case under investigation. 

Let $n$ be the number of lattice points in the
interval $[x_1,L]$. We consider the scaled 
random variable 
$\tilde{Z}^{{\rm O}}=Z^{{\rm O}}/n^{1/\alpha}$ with 
$\alpha=T/T_g$ and $T<T_g$.
The Laplace $z \to u$ transform of the PDF of $\tilde{Z}^{{\rm O}}$ is
found using Eq. 
(\ref{eq04a}) and $\rho(E)$
\begin{equation}
\hat{G}_{\tilde{Z}^{{\rm O}}}\left(u\right) = \exp \left\{ \sum_{x=x_1}^L \ln \left[ \hat{\psi} \left( { u e^{ - {U^{{\rm det}}_x \over T }} \over n^{1/\alpha} } \right) \right] \right\}
\label{eq06}
\end{equation}
where $\hat{\psi}(u) = \int_0 ^\infty \exp( - u \tau) \psi(\tau) {\rm d} \tau$.
We now consider the limit of large $n$. We use the small $u$ expansion
\begin{equation}
\ln\left[ \hat{\psi}(u) \right] \sim - A u^\alpha + {\alpha \over 1 - \alpha} u + \cdots
\label{eq07}
\end{equation}
where $A=\alpha|\Gamma(-\alpha)|$  
and from Eqs. (\ref{eq06},\ref{eq07}) we find
$$ \hat{G}_{\tilde{Z}^{{\rm O}} } \left( u \right) \sim  $$
\begin{equation}
\exp \left\{
-  {A u^\alpha \over n} \sum_{x=x_1} ^L e^{ -{U_x ^{{\rm det}} \alpha \over T}}  +  {\alpha u \over \left(1 - \alpha\right) n^{1/\alpha} }  \sum_{x=x_1} ^L e^{ - {U^{{\rm det} } _x \over T}} + \cdots  \right\}.
\label{eqBP07}
\end{equation}
In the continuum limit of $a\to 0$ $n\to \infty$ and $L-x_1=a n$ 
remaining finite we may replace the summation with integration
and find the stretched exponential 
\begin{equation}
\hat{G}_{\tilde{Z}^{{\rm O}}} \left( u \right) \sim 
\exp \left[ - A   { \int_{x_1}^L e^{- {U^{{\rm det}}(x) \over T_g }}  {\rm d} x  \over L-x_1} u^{\alpha} \right], 
\label{eqBP09}
\end{equation}
where $U^{{\rm det}}(x)$ is the deterministic field in the continuum limit.
The inverse Laplace transform of Eq. (\ref{eqBP09}) is the
one sided L\'evy stable law. 

 A similar calculation is made for $Z^{{\rm NO}}$. We invert
the Laplace transform Eq. (\ref{eqBP09}), switch back to the original variable
$Z^{{\rm O}}$ instead of the scaled one $\tilde{Z}^{{\rm O}}$, and 
find using Eq. 
(\ref{eq05})
\begin{equation}
f\left( \overline{p} \right) \sim  { 1 \over \left( {\cal R} \right)^{1/\alpha} } \int_0 ^\infty {\rm d} z z l_{\alpha} \left[ \left( 1 - \overline{p} \right) z \right] l_{\alpha} \left[ { \overline{p} z \over \left( {\cal R} \right)^{1/\alpha} } \right]
\label{eq100}
\end{equation}
with
\begin{equation}
{\cal R} = {P_B \left( T_g \right) \over 1 - P_B \left(T_g\right) }
\label{eqBP17a}
\end{equation}
In Eq.  (\ref{eq100}) $l_{\alpha}(z)$ is the one sided L\'evy stable PDF
whose Laplace pair is $\hat{l}_{\alpha}(u)\equiv \exp(- u^\alpha)$. 
$P_B(T_g)$ is Boltzmann's probability of finding the
particle in the observation domain calculated using the deterministic
field with a temperature $T_g$  
\begin{equation}
P_B(T_g) = {\int_{x_1}^L \exp(- U^{{\rm det}}(x)/T_g) {\rm d} x \over  Z(T_g)}.
\label{eq109}
\end{equation}
Solving the integral Eq. (\ref{eq100}) 
we find
the Lamperti \cite{Lamp} PDF 
\begin{equation}
f\left( \overline{p} \right)  \sim
{ \sin \pi \alpha \over \pi }
{{\cal R} \overline{p} ^{\alpha -1} \left( 1 - \overline{p} \right)^{\alpha-1} \over
{\cal R}^2 \left( 1 - \overline{p} \right)^{2 \alpha} + \overline{p} ^{2 \alpha} + 2
{\cal R} \left( 1 -  \overline{p} \right)^{\alpha} \overline{p} ^\alpha \cos \pi \alpha }.
\label{eq101}
\end{equation}
Eq. 
(\ref{eqBP17a},
\ref{eq101}) 
are the main results of this manuscript, soon to be discussed in detail,
which are
valid in the glassy phase $T<T_g$.

\begin{figure}
\begin{center}
\includegraphics[width=.4\textwidth,angle=-90]{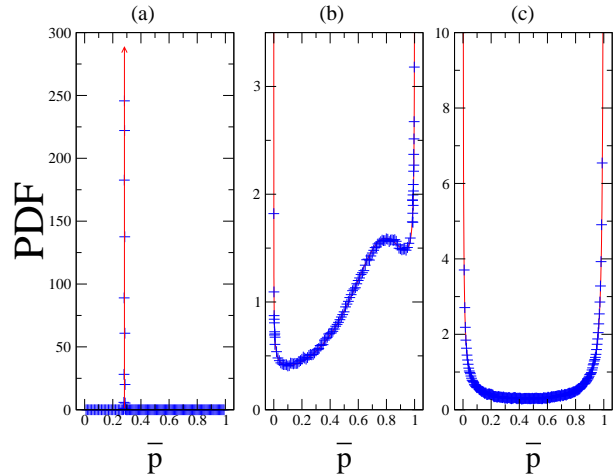}
\end{center}
\caption{
The PDF of the occupation fraction 
for the deterministic field $U(x) = {\cal F} x$. 
{\bf (a)} For $T= 3 T_g$  we find a delta function
centered on the value given by Boltzmann's statistics.  When
$T/T_g =0.7$ [panel {\bf (b)}] 
the non trivial distribution of the occupation
fraction has three peaks  while for $T/T_g =0.3$ [panel {\bf (c)}]
 the distribution
is bi-modal.
The $+$ are simulations and the curve is 
the theoretical prediction Eq. (\protect{\ref{eq101}})
without fitting. 
We used ${\cal F}=1,T_g = 1,a=10^{-5}$ 
and the observation domain $0<x<1$. 
}
\label{fig2}
\end{figure}
%

For $T>T_g$ and in the same limit we have
the usual canonical behavior
\begin{equation}
f\left( \overline{p} \right) \sim \delta \left( \overline{p} - P_B(T) \right). 
\label{eq102}
\end{equation}
Eq. (\ref{eq102})
shows that when $T>T_g$ the disorder plays no role, 
indicating the reproducibility of Boltzmann's
statistics Eq. 
(\ref{eqInt01}),
when the disorder is weak. 

We now 
discuss the behaviors found in our main  Eqs. 
(\ref{eqBP17a},
\ref{eq101}). These equations give the distribution of the
occupation times, which is the generalization of the
usual Boltzmann law Eq. (\ref{eq102}).  
 The parameter ${\cal R}$ is called the asymmetry parameter, and if 
${\cal  R}=1$, $f(\overline{p})$ is symmetric, for example
when $T/T_g =1/2$ and ${\cal R} = 1$ we get the 
arcsine PDF.  
The asymmetry parameter ${\cal R}$  is calculated by the usual type in integral
over the
Boltzmann factor, {\em however now the temperature $T_g$ is
the relevant temperature} not $T$ [see Eq. (\ref{eqBP17a})]. Roughly
speaking there are two sources of fluctuations:
the disorder characterized
by  $T_g$,
and the temperature $T$. Hence when $T<T_g$ the relevant temperature
is the ``temperature of the disorder'' that is $T_g$. For example
using Eq. (\ref{eqBP17a},\ref{eq101},\ref{eq102}) 
the average occupation fraction has the following surprising
 discontinuous
behavior, 
\begin{equation}
\langle \overline{p} \rangle = \left\{
\begin{array}{c c}
P_B(T_g) \ & T<T_g \\
P_B(T) \ & T>T_g.
\end{array}
\right.
\label{eq303}
\end{equation}
The average occupation fraction freezes in the colder glassy phase
of $T<T_g$ in the sense that it does not depend on the temperature $T$,
for any type of deterministic  binding field. 

In Fig. \ref{fig2} 
we demonstrate our results comparing our theory with
numerical simulations on a lattice.
We consider the situation where the deterministic field
is $U^{{\rm det}}(x) =  {\cal F} x$ and $0<x$, and
the observation domain is $0<x< T_g/{\cal F}$.
In Fig. \ref{fig2}(a) with $T>T_g$  we see that  the distribution
of  occupation 
fraction  is  very narrow
with
%
$\overline{p} = P_B(T=3 T_g)=1-e^{-1/3}$
%
indicating that the disorder is not important. 
In contrast when $T<T_g$ the behavior of the occupation fraction
changes dramatically  and
$\overline{p}$ is non self averaging and random.

 Eq. (\ref{eq101}) shows that when $T/T_g\ll 1$ the PDF of occupation
fraction  is essentially composed of two delta functions
centered on $\overline{p} =1$ and $\overline{p}=0$.
Namely for some samples of disorder the particle is within
the observation zone during all the observation time
$t$ ($\overline{p} =1$) and  in other samples the particle
is  never in the observation
zone ($\overline{p} = 0$). 
This behavior is easy to understand
when $T\to 0$ the minimum of the random potential energy is the most
populated, and this minimum can be found either in the observation
zone or out of it. As shown in Fig. \ref{fig2} (c),
 for small but finite $T$ we  have a non-trivial
bi-modal $U$ shape of the PDF,  which reflects this low temperature behavior. 
As the temperature increases we start seeing a third peak in the PDF
of the occupation fraction being developed 
close to the ensemble average [see center peak in Fig. \ref{fig2} (b)], 
so when $T \to T_g$ the self-averaging 
phase is approached.   

\begin{figure}
\begin{center}
\includegraphics[width=.4\textwidth]{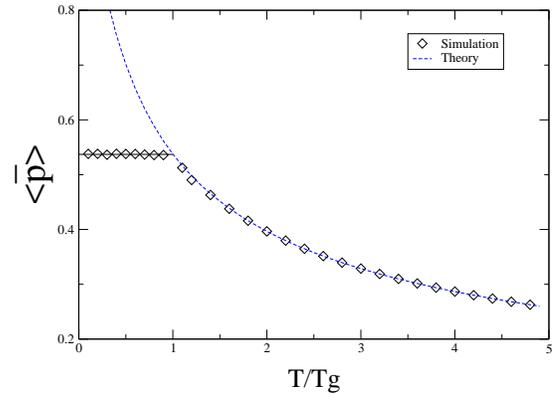}
\end{center}
\caption{
The averaged  occupation fraction versus $T/T_g$. 
When $T>T_g$, $\langle \overline{p} \rangle = P_B(T)$ 
namely usual Boltzmann
theory applies,  
while for
$T<T_g$, 
$\langle \overline{p} \rangle = P_B(T_g)$ 
which is independent of the temperature $T$. 
The lines are theoretical predictions Eq. (\protect{\ref{eq303}}) 
and the diamonds are
simulation results with no fitting.  
}
\label{fig4}
\end{figure}

 In Fig. \ref{fig4}  we show the averaged occupation fraction versus
temperature $T$ using the same
deterministic potential field as in Fig. \ref{fig2}. 
For $T>T_g$ the average occupation
fraction is $P_B(T)$  and hence as the temperature 
is decreased the average occupation time increases,
since 
the particles condensate closer to the minimum of the deterministic
field which is on  $x=0$ as the temperature is reduced.  
However when $T=T_g$ we see in Fig. \ref{fig4} 
a type of phase transition in the behavior
of the averaged occupation fraction,  
 and  it does
not depend on the temperature when $T<T_g$, as predicted
by our theory Eq. (\ref{eq303}). 
 We note that the convergence of the 
numerical results to the exact ones derived in this paper turned
out to be relatively slow close to the temperature $T=T_g$
if compared with temperatures far from $T_g$.

 Finally let us discuss the generality of our results beyond
the quenched trap model. We have divided our system into
two, the observation domain and the rest of the system. The
partition functions of these domains are random variables, due to the
randomness of the underlying Hamiltonian. As we showed,  one key ingredient
of our theory is that the PDFs of these two partition functions
are one sided L\'evy stable PDFs. In that case we proved that 
our main Eq. (\ref{eq101})  describes the statistics of occupation time.
We know that L\'evy statistics is common in Physics and
well established mathematically,  and
hence it seems to us
natural to expect that partition functions of random systems
may have a L\'evy
distribution.  Indeed a partition function is a sum
over energy states and if these states are random 
the connection of the distribution of partition functions
with limit theorems of sums of random variables is expected
to be general.
In particular we
could show that our main results describe also occupation
time  statistics in several other models of quenched disorder:
random comb model which is a simple model of a random
walk on a loop-less random fractal \cite{Havlin}, 
or models of anomalous diffusion of a particle
on structures with distributed dangling bonds 
in the presence of bias \cite{RevBouchaud,Bar83,Bun86}.  
Unlike the trap model,  
in these models geometry  is the main factor
responsible for the non-trivial occupation fraction.   
Sure the exponent does not generally turn out to be $\alpha=T/T_g$ as
in the quenched trap model, but our main results
are valid as we will discuss in a longer publication. 

{\bf Acknowledgment} This work was supported by the 
Israel Science Foundation. EB thanks A. Comtet, S. Majumdar,
and  G. Margolin for
discussions.

\end{document}